\documentclass[12pt,prd,aps,tightenlines,nofootinbib,longbibliography]{revtex4-2}
\usepackage{bm}
\usepackage{graphics}
\usepackage{rotating}
\usepackage{epsfig}
\usepackage{multirow}
\begin{document}
\title{Masses of the $QQ\bar Q\bar Q$ tetraquarks in the
relativistic diquark--antidiquark picture}
\author{R.~N.~Faustov}
\author{V.~O.~Galkin}
\affiliation{Institute of Cybernetics and Informatics in Education,
  Federal Research Center ``Computer Science and Control'', Russian Academy of Sciences, 
  Vavilov Street 40, 119333 Moscow, Russia}
\author{E. M. Savchenko}
\affiliation{Faculty of Physics, M.V. Lomonosov Moscow State University,  119991 Moscow, Russia}
\begin{abstract}
Masses of the ground state teraquarks composed from heavy $c$ and $b$
quarks and antiquarks are calculated in the diquark-antidiquark
picture in the framework of the relativistic quark model based on the
quasipotential approach. The quasipotentials of the quark-quark and
diquark-antidiquark interactions are constructed similarly to the
previous consideration of mesons and  baryons. It is assumed that the diquark and
antidiquark interact in the tetraquark as a whole and the internal
structure of the diquarks is taken into account.
All such tetraquarks  are found above the thresholds of
decays to two heavy quarkonia. This is a result of the consideration of the
diquark  not to be a point-like object. Therefore such tetraquarks can be observed as broad
structures decaying dominantly to quarkonia. The broad structure next
to the di-$J/\psi$ mass threshold, recently observed by the LHCb
Collaboration, can correspond to the ground
$2^{++}$-state  tetraquark consisting of four charm quarks.

\end{abstract}

\maketitle

\section{Introduction}
\label{sec:intr}

Theoretical and experimental investigations of the properties of
exotic hadrons have attracted
substantial interest, especially  in last two decades. This subject
became a hot topic since the first explicit experimental evidence of the
existence of hadrons with compositions different from usual $q\bar q$
for mesons and $qqq$ for baryons became available (for recent reviews, see \cite{Liu:2019zoy,Brambilla:2019esw,Yang:2020atz} and
references therein).  Candidates for both the exotic tetraquark
$qq\bar q\bar q$ and pentaquark
$qqqq\bar q$ states were found. However, in the literature there is no
consensus about the composition of these states \cite{Liu:2019zoy,Brambilla:2019esw,Yang:2020atz}. For example,
significantly different interpretations for the
$qq\bar q\bar q$ candidates were proposed: molecules
composed from two mesons loosely bound by the meson exchange,
compact  tetraquarks composed from a diquark and antidiquark bound by
strong forces,
hadroquarkonia composed of a heavy quarkonium embedded in a light meson, kinematic cusps, etc. Discrimination between different
approaches is a very complicated experimental task.

The investigation of exotic $QQ\bar Q\bar Q$ states  consisting of heavy
quarks ($Q=c$ and/or $b$) is of special interest, since their nature
can be determined more easily. They should be predominantly compact
tetraquarks. Indeed, a molecular configuration is 
unlikely. Only heavy $Q\bar Q$ mesons can be exchanged between
constituents in such a molecule, and the arising Yukawa-type potential
is not strong enough to provide binding. Soft gluons can be exchanged
between two heavy quarkonia, leading to the so-called QCD van der
Waals force. Such a force is known to be attractive, though whether it
is strong enough to form a bound state remains unclear. The hadroquarkonium picture
is not applicable. Thus, the diquark
($QQ$)-antidiquark ($\bar Q\bar Q$) configuration is preferable.

The CMS \cite{cms} and LHCb \cite{lhcb} Collaborations
searched for the tetraquark states composed only of bottom quarks
in the $\Upsilon$-pair production. No evidence of such states was
found.  Very recently, the LHCb Collaboration \cite{Aaij:2020fnh} reported results of the
study of the $J/\psi$-pair invariant mass spectrum in proton-proton
collision data at the center-of-mass energies of $\sqrt{s}=7.8$ and 13
TeV.  A narrow structure around 6.9 GeV and a broad structure just
above twice the $J/\psi$ mass were observed. This discovery caused
considerable theoretical activity interpreting these data (see
\cite{Maiani:2020pur,Richard:2020hdw,Karliner:2020dta,Dong:2020nwy}
and references therein).

In this paper we calculate masses of the ground-state $QQ\bar
Q\bar Q$ tetraquarks in the framework of the relativistic quark model
based on the quasipotential approach. It is assumed that such
tetraquarks are composed from the doubly-heavy diquark ($QQ$) and
antidiquark ($\bar Q\bar Q$). Such approximation significantly
simplifies calculations, since instead of the very complicated
relativistic four-body problem we need to solve two more simple
relativistic two-body problems. First, masses and wave functions of
diquarks (antidiquarks) are obtained by solving the relativistic
quark-quark (antiquark-antiquark)
quasipotential equation. Second, masses of tetraquarks are calculated
by considering them to be the diquark-antidiquark bound states. The
quasipotentials  of the corresponding interactions are constructed
using the same assumptions about their structure and parameters which
were previously used for the investigation of the different properties
of mesons and baryons \cite{egf,efgm,qqregge,barregge,Faustov:2015eba}.  The
spin-independent   and spin-dependent relativistic contributions to the
quasipotentials of the $QQ$ interaction in a diquark $d$ and the
$d\bar d$ interaction in a tetraquark
are considered nonperturbatively. It is assumed that a diquark and antidiquark in a
tetraquark interact as a whole: thus  interactions between quarks from
a diquark with antiquarks from an antidiquark are not considered. It
is important to point out that diquarks and antidiquarks are not the
point-like objects. Their short-distance interaction with gluons is
smeared by the form factors which are calculated in terms of the
overlap integrals of the diquark wave functions. 
Such an approach was previously applied for the calculation of the masses
of heavy ($qQ\bar q\bar Q$, $QQ\bar q\bar q$, $Qq\bar q\bar q$) and
light ($qq\bar q\bar q$) tetraquarks
\cite{Ebert:2005nc,extetr,efgl,Ebert:2008id,Ebert:2010af}.

This paper is organized as follows. In Sec.~\ref{sec:rqm} we describe
our relativistic quark model. The quasipotentials of the $QQ$
and $d\bar d$ interactions are presented. The masses of the
doubly-heavy diquarks and the form factors of their interaction with
gluons are obtained. In Sec.~\ref{sec:mht}, the masses of the $QQ\bar
Q\bar Q$ tetraquarks are calculated. They are confronted with the lowest
thresholds for the fall-apart decays to two heavy quarkonia. Detailed
comparisons with previous theoretical predictions within different
approaches are given. Finally, we present our conclusions and summary of the
obtained results in Sec.~\ref{sec:concl}.

\section{Relativistic diquark-antidiquark model}
\label{sec:rqm}

For the calculation of the masses of tetraquarks, we use the relativistic quark model based on the quasipotential approach and the diquark-antidiquark picture of tetraquarks. First, we calculate the masses  and  wave functions ($\Psi_{d}$) of the doubly-heavy diquarks as the bound quark-quark states.  Second,  the masses of the tetraquarks  and their wave functions ($\Psi_{T}$) are obtained for the bound diquark-antidiquark states. These wave functions are solutions of the  Schr\"odinger-type quasipotential equations \cite{efgl}
\begin{equation}
\label{quas}
{\left(\frac{b^2(M)}{2\mu_{R}}-\frac{{\bf
p}^2}{2\mu_{R}}\right)\Psi_{d,T}({\bf p})} =\int\frac{d^3 q}{(2\pi)^3}
 V({\bf p,q};M)\Psi_{d,T}({\bf q}),
\end{equation}
with the  on-mass-shell relative momentum squared  given by
\begin{equation}
{b^2(M) }
=\frac{[M^2-(m_1+m_2)^2][M^2-(m_1-m_2)^2]}{4M^2},
\end{equation}
and the relativistic reduced mass
\begin{equation}
\mu_{R}=\frac{E_1E_2}{E_1+E_2}=\frac{M^4-(m^2_1-m^2_2)^2}{4M^3}.
\end{equation}
The  on-mass-shell energies  $E_1$, $E_2$ are defined as follows:
\begin{equation}
\label{ee}
E_1=\frac{M^2-m_2^2+m_1^2}{2M}, \quad E_2=\frac{M^2-m_1^2+m_2^2}{2M}.
\end{equation}
The bound-state masses of a diquark or a tetraquark are $M=E_1+E_2$ , where
$m_{1,2}$ are the masses of quarks ($Q_1$ and $Q_2$) which form
the diquark, or of the diquark ($d$) and antidiquark ($\bar d'$) which
form the heavy tetraquark ($T$), while ${\bf p}$ is their relative
momentum. 

The quasipotential operator $V({\bf p,q};M)$ in Eq.~(\ref{quas}) is  constructed with the help of the off-mass-shell scattering amplitude, projected onto the positive-energy states.
The quark-quark  ($QQ'$) interaction quasipotential \footnote{We
  consider diquarks in a tetraquark, as in a baryon, to be in the color triplet state, since in the
  color sextet there is a repulsion between two quarks.} is considered to be 1/2 of the quark-antiquark   ($Q\bar Q'$) interaction and is given by \cite{egf}
 \begin{equation}
\label{qpot}
V({\bf p,q};M)=\bar{u}_{1}(p)\bar{u}_{2}(-p){\cal V}({\bf p}, {\bf
q};M)u_{1}(q)u_{2}(-q),
\end{equation}
with
\[
{\cal V}({\bf p,q};M)=\frac12\left[\frac43\alpha_sD_{ \mu\nu}({\bf
k})\gamma_1^{\mu}\gamma_2^{\nu}+ V^V_{\rm conf}({\bf k})
\Gamma_1^{\mu}({\bf k})\Gamma_{2;\mu}(-{\bf k})+
 V^S_{\rm conf}({\bf k})\right].
\]
Here,  $D_{\mu\nu}$ is the
gluon propagator in the Coulomb gauge, $u(p)$ are the Dirac spinors and $\alpha_s$ is the running QCD coupling constant with freezing
\begin{equation}
  \label{eq:alpha}
  \alpha_s(\mu^2)=\frac{4\pi}{\displaystyle\left(11-\frac23n_f\right)
\ln\frac{\mu^2+M_B^2}{\Lambda^2}}, 
\end{equation}
where the scale $\mu$ is chosen to be equal to $2m_1 m_2/(m_1+m_2)$, the background mass is $M_B=2.24\sqrt{A}=0.95$~GeV, and $n_f$ is the number of flavours.
 The effective long-range vector vertex  contains both Dirac and Pauli terms
\cite{egf} 
\begin{equation}
\Gamma_{\mu}({\bf k})=\gamma_{\mu}+
\frac{i\kappa}{2m}\sigma_{\mu\nu}\tilde k^{\nu}, \qquad \tilde
k=(0,{\bf k}),
\end{equation}
where $\kappa$ is the long-range anomalous chromomagnetic moment. In the nonrelativistic limit the vector and scalar confining potentials in configuration space have the form
\begin{eqnarray}
V^V_{\rm conf}(r)&=&(1-\varepsilon)(Ar+B),\qquad V^S_{\rm conf}(r)=\varepsilon (Ar+B),\nonumber\\[1ex]
V_{\rm conf}(r)&=&V^V_{\rm conf}(r)+V^S_{\rm conf}(r)=Ar+B,
\end{eqnarray}
where $\varepsilon$ is the mixing coefficient.  Therefore in the nonrelativistic limit the $QQ'$ quasipotential reduces to
\begin{equation}
V^{\rm NR}_{QQ'}(r)=\frac12V^{\rm NR}_{Q\bar Q'}(r)=\frac12\left(-\frac43\frac{\alpha_s}{r}+Ar+B\right),
\end{equation} 
reproducing the usual Cornel potential. Thus, our quasipotential can be
viewed as its relativistic generalization.  It contains both
spin-independent and spin-dependent relativistic contributions. 

Constructing the diquark-antidiquark ($d\bar d'$) 
quasipotential, we use the same assumptions about the structure of the
short- and long-range interactions. Taking into account the integer
spin of a diquark in the color triplet state, the quasipotential is given by \cite{efgl,extetr}
\begin{eqnarray}
\label{dpot} 
V({\bf p,q};M)&=&\frac{\langle
d(P)|J_{\mu}|d(Q)\rangle} {2\sqrt{E_dE_d}} \frac43\alpha_sD^{
\mu\nu}({\bf k})\frac{\langle d'(P')|J_{\nu}|d'(Q')\rangle}
{2\sqrt{E_{d'}E_{d'}}}\nonumber\\[1ex]
&&+\psi^*_d(P)\psi^*_{d'}(P')\left[J_{d;\mu}J_{d'}^{\mu} V_{\rm
conf}^V({\bf k})+V^S_{\rm conf}({\bf
k})\right]\psi_d(Q)\psi_{d'}(Q'),
\end{eqnarray}
where $\psi_d(p)$  and $J_{d;\mu}$ are the wave function and  effective long-range vector
vertex of the diquark, respectively.   The vertex of the diquark-gluon
interaction $\langle d(P)|J_{\mu}|d(Q)\rangle$  accounts for the internal structure of  
the diquark and leads to emergence of the form factor $F(r)$ smearing the one-gluon exchange potential. 

All parameters of the model were fixed previously
\cite{egf,efgm,qqregge,barregge} from the consideration of meson and
baryon properties. They are as follows: The constituent heavy quark
masses are $m_b=4.88$ GeV, $m_c=1.55$ GeV. The parameters of the
quasipotential are $A=0.18$ GeV$^2$, $B=-0.3$~GeV, $\Lambda=413$~MeV. The mixing coefficient of
vector and scalar confining potentials $\varepsilon=-1$ and the universal
Pauli interaction constant $\kappa=-1$.

The resulting diquark-antidiquark potential for the tetraquark ground
states (the orbital momentum $L=0$), where quark energies
$\epsilon_{1,2}(p)$ were replaced by the on-shell energies $E_{1,2}$
(\ref{ee}) to remove the non-locality, is given by  \cite{efgl}:
\begin{eqnarray}
 \label{eq:pot}
 V(r)&=& \hat V_{\rm Coul}(r)+V_{\rm conf}(r)+\frac1{E_1E_2}\Biggl\{{\bf
 p}\left[\hat V_{\rm Coul}(r)+V^V_{\rm conf}(r)\right]{\bf p} -\frac14
\Delta V^V_{\rm conf}(r)\nonumber\\[1ex]
 &&
+\frac23\Delta \hat V_{\rm
Coul}(r){\bf S}_1\cdot{\bf S}_2\Biggr\}.
\end{eqnarray}
Here $$\hat V_{\rm Coul}(r)=-\frac{4}{3}\alpha_s
\frac{F_1(r)F_2(r)}{r}$$ is the Coulomb-like one-gluon exchange
potential which takes into account the finite sizes of the diquark
and antidiquark through corresponding form factors $F_{1,2}(r)$. ${\bf
  S}_{1,2}$ are the diquark and antidiquark spins. The numerical analysis shows that this form factor can be approximated with high accuracy by the expression
\begin{equation}
 \label{eq:fr}
 F(r)=1-e^{-\xi r -\zeta r^2}.
\end{equation}
Such a form factor smears the one-gluon exchange potential and removes spurious singularities in the local relativistic quasipotential, thus allowing one to use it nonperturbatively to find the numerical solution of the quasipotential equation. The masses and parameters of heavy diquarks are the same as in the doubly heavy baryons \cite{efgm} and are given in Table~\ref{tab:dqm}.

\begin{table}
 \caption{Masses $M$ and form factor parameters of heavy $QQ'$ diquarks. $S$ and $A$
 denote scalar and axial-vector diquarks, antisymmetric $[Q,Q']$ and
 symmetric $\{Q,Q'\}$ in flavour, respectively. }
 \label{tab:dqm}
\begin{ruledtabular}
\begin{tabular}{cccccccc}
Quark& Diquark&
\multicolumn{3}{l}{\underline{\hspace{2.5cm}$Q=c$\hspace{2.5cm}}}
\hspace{-3.4cm}
&\multicolumn{3}{l}{\underline{\hspace{2.5cm}$Q=b$\hspace{2.5cm}}}
\hspace{-3.4cm} \\ content &type & $M$ (MeV)&$\xi$ (GeV)&$\zeta$
(GeV$^2$) & $M$ (MeV)&$\xi$ (GeV)&$\zeta$ (GeV$^2$) \\ \hline
 $[Q,c]$ & $S$& & & & 6519 & 1.50
&0.59\\ $\{Q,c\}$& $A$& 3226& 1.30& 0.42 & 6526 & 1.50 &0.59\\
$\{Q,b\}$& $A$& 6526 & 1.50 &0.59& 9778 & 1.30 &1.60
 \end{tabular}
\end{ruledtabular}
\end{table}

\section{Masses of $QQ\bar Q\bar Q$ tetraquarks}
\label{sec:mht}

\begin{table}
 \caption{Masses $M$ of the neutral heavy diquark ($QQ'$)-antidiquark ($\bar Q\bar Q'$)
 states. $T$ is the threshold for the decays into two
 heavy-($Q\bar Q'$) mesons and $\Delta=M-T$. All values are
 given~in~MeV.}
 \label{tab:QQmass}
\begin{ruledtabular}
\begin{tabular}{  c c c c c c c }
Composition &$d\bar d$  & $J^{PC}$ & $M$ & Threshold& $T$ & $\Delta$
\\
\hline
 \multirow{4}{*}{$cc \bar c \bar c$} & \multirow{4}{*}{$A \bar A$} & \multirow{2}{*}{$0^{++}$} & \multirow{2}{*}{6190} & $\eta_{c}(1S)\eta_{c}(1S)$ & 5968 & 222
\\
\cline{5-7}
  & & & &  $J/\psi(1S)J/\psi(1S)$ & 6194 & -4
\\
\cline{3-7}
  & & $1^{+-}$ & 6271 & $\eta_{c}(1S)J/\psi(1S)$ & 6081 & 190
\\
\cline{3-7}
  & & $2^{++}$ & 6367 & $J/\psi(1S)J/\psi(1S)$ & 6194 & 173
\\
\cline{1-7}

  \multirow{21}{*}{$cb \bar c \bar b$} & \multirow{10}{*}{$A \bar A$} & \multirow{4}{*}{$0^{++}$} & \multirow{4}{*}{12813} & $\eta_{c}(1S)\eta_{b}(1S)$ & 12383 & 430
\\
\cline{5-7}
  & & & & $J/\psi(1S)\Upsilon(1S)$ & 12557 & 256
\\
\cline{5-7}
  & & & & $B_{c}^{\pm}  B_{c}^{\mp}$ & 12550 & 263
\\
\cline{5-7}
  & & & & $B_{c}^{*\pm} B_{c}^{*\mp}$ & 12666 & 147
\\
\cline{3-7}
  & & \multirow{4}{*}{$1^{+-}$} & \multirow{4}{*}{12826} & $\eta_{c}(1S)\Upsilon(1S)$ & 12444 & 382
\\
\cline{5-7}
  & & & & $J/\psi(1S)\eta_{b}(1S)$ & 12496 & 330
\\
\cline{5-7}
  & & & & $B_{c}^{\pm}  B_{c}^{*\mp}$ & 12608 & 218
\\
\cline{5-7}
  & & & & $B_{c}^{* \pm}  B_{c}^{* \mp}$ & 12666 & 160
\\
\cline{3-7}
  & & \multirow{2}{*}{$2^{++}$} & \multirow{2}{*}{12849} & $J/\psi(1S)\Upsilon(1S)$ & 12557 & 292
\\
\cline{5-7}
  & & & & $B_{c}^{* \pm}  B_{c}^{*\mp}$ & 12666 & 183
\\
\cline{2-7}
  & \multirow{7}{*}{$\frac{1}{\sqrt{2}}(A \bar S \pm S \bar A)$} & \multirow{3}{*}{$1^{++}$} & \multirow{4}{*}{12831} & $J/\psi(1S)\Upsilon(1S)$ & 12557 & 274
\\
\cline{5-7}
  & & & & $B_{c}^{\pm} B_{c}^{*\mp}$ & 12608 & 223
\\
\cline{5-7}
  & & & & $B_{c}^{*\pm}  B_{c}^{*\mp}$ & 12666 & 165
\\
\cline{3-7}
  & & \multirow{4}{*}{$1^{+-}$} &  \multirow{4}{*}{12831} & $\eta_{c}(1S)\Upsilon(1S)$ & 12444 & 387
\\
\cline{5-7}
  & & & & $J/\psi(1S)\eta_{b}(1S)$ & 12496 & 335
\\
\cline{5-7}
  & & & & $B_{c}^{\pm} B_{c}^{*\mp}$ & 12608 & 223
\\
\cline{5-7}
  & & & & $B_{c}^{*\pm}  B_{c}^{*\mp}$ & 12666 & 165
\\
\cline{2-7}
  & \multirow{4}{*}{$S \bar S$} & \multirow{4}{*}{$0^{++}$} & \multirow{4}{*}{12824} & $\eta_{c}(1S)\eta_{b}(1S)$ & 12383 & 441
\\
\cline{5-7}
  & & & & $J/\psi(1S)\Upsilon(1S)$ & 12557 & 267
\\
\cline{5-7}
  & & & & $B_{c}^{\pm}  B_{c}^{\mp}$ & 12550 & 274
\\
\cline{5-7}
  & & & & $B_{c}^{*\pm}  B_{c}^{*\mp}$ & 12666 & 158
\\
\cline{1-7}
  \multirow{4}{*}{$bb \bar b \bar b$} & \multirow{4}{*}{$A \bar A$} & \multirow{2}{*}{$0^{++}$} & \multirow{2}{*}{19314} & $\eta_{b}(1S)\eta_{b}(1S)$ & 18797 & 517
\\
\cline{5-7}
  & & & &  $\Upsilon(1S)\Upsilon(1S)$ & 18920 & 394
\\
\cline{3-7}
  & & $1^{+-}$ & 19320 & $\eta_{b}(1S)\Upsilon(1S)$ & 18859 & 461
\\
\cline{3-7}
  & & $2^{++}$ & 19330 & $\Upsilon(1S)\Upsilon(1S)$ & 18920 & 410
\\
\end{tabular}
\end{ruledtabular}
\end{table}

\begin{table}
 \caption{Masses $M$ of the charged heavy diquark--antidiquark 
 states. $T$ is the threshold for the decays into two
 heavy ($Q\bar Q'$) mesons and $\Delta=M-T$. All values are
 given~in~MeV.}
 \label{tab:QQQQ'mass}
\begin{ruledtabular}
\begin{tabular}{  c c c c c c c }
Composition &$d\bar d$  & $J^{P}$ & $M$ & Threshold& $T$ & $\Delta$
\\
\hline
 \multirow{9}{*}{$cc \bar c \bar b, cb \bar c \bar c$} & \multirow{6}{*}{$A \bar A$} & \multirow{2}{*}{$0^{+}$} & \multirow{2}{*}{9572} & $\eta_{c}(1S)B_{c}^{\pm}$ & 9259 & 313
\\
\cline{5-7}
  & & & &  $J/\psi(1S)B_{c}^{* \pm}$ & 9430 & 142
\\
\cline{3-7}
  & & \multirow{3}{*}{$1^{+}$} & \multirow{3}{*}{9602} & $\eta_{c}(1S)B_{c}^{* \pm}$ & 9317 & 285
\\
\cline{5-7}
  & & & & $J/\psi(1S)B_{c}^{\pm}$ & 9372 & 230
\\
\cline{5-7}
  & & & & $J/\psi(1S)B_{c}^{* \pm}$ & 9430 & 172
\\
\cline{3-7}
  & & $2^{+}$ & 9647 & $J/\psi(1S)B_{c}^{* \pm}$ & 9430 & 217
\\
\cline{2-7}
  & \multirow{3}{*}{$A \bar S$,  $ S \bar A$} & \multirow{3}{*}{$1^{+}$} & \multirow{3}{*}{9619} & $\eta_{c}(1S)B_{c}^{* \pm}$ & 9317 & 302
\\
\cline{5-7}
  & & & & $J/\psi(1S)B_{c}^{\pm}$ & 9372 & 247
\\
\cline{5-7}
  & & & & $J/\psi(1S)B_{c}^{* \pm}$ & 9430 & 189
\\
\cline{1-7}
  \multirow{5}{*}{$cc \bar b \bar b, bb \bar c \bar c$} & \multirow{5}{*}{$A \bar A$} & \multirow{2}{*}{$0^{+}$} & \multirow{2}{*}{12846} & $B_{c}^{\pm}B_{c}^{\pm}$ & 12550 & 296
\\
\cline{5-7}
  & & & & $B_{c}^{* \pm}B_{c}^{* \pm}$ & 12666 & 180
\\
\cline{3-7}
  & & \multirow{2}{*}{$1^{+}$} & \multirow{2}{*}{12859} & $B_{c}^{\pm} B_{c}^{* \pm}$ & 12608 & 251
\\
\cline{5-7}
  & & & & $B_{c}^{* \pm}B_{c}^{* \pm}$ & 12666 & 193
\\
\cline{3-7}
  & & \multirow{1}{*}{$2^{+}$} & \multirow{1}{*}{12883} & $B_{c}^{* \pm}B_{c}^{* \pm}$ & 12666 & 217
\\
\cline{1-7}
  \multirow{9}{*}{$cb \bar b \bar b, bb \bar c \bar b$} & \multirow{6}{*}{$A \bar A$} & \multirow{2}{*}{$0^{+}$} & \multirow{2}{*}{16109} & $B_{c}^{\pm}\eta_{b}(1S)$ & 15674 & 435
\\
\cline{5-7}
  & & & &  $B_{c}^{* \pm}\Upsilon(1S)$ & 15793 & 316
\\
\cline{3-7}
  & & \multirow{3}{*}{$1^{+}$} & \multirow{3}{*}{16117} & $B_{c}^{\pm}\Upsilon(1S)$ & 15735 & 382
\\
\cline{5-7}
  & & & & $B_{c}^{* \pm}\eta_{b}(1S)$ & 15732 & 385
\\
\cline{5-7}
  & & & & $B_{c}^{* \pm}\Upsilon(1S)$ & 15793 & 324
\\
\cline{3-7}
  & & $2^{+}$ & 16132 & $B_{c}^{* \pm}\Upsilon(1S)$ & 15793 & 339
\\
\cline{2-7}
  & \multirow{3}{*}{$S \bar A$, $A \bar S  $} & \multirow{3}{*}{$1^{+}$} & \multirow{3}{*}{16117} & $B_{c}^{\pm}\Upsilon(1S)$ & 15735 & 382
\\
\cline{5-7}
 & & & & $B_{c}^{* \pm}\eta_{b}(1S)$ & 15732 & 385
\\
\cline{5-7}
  & & & & $B_{c}^{* \pm}\Upsilon(1S)$ & 15793 & 324
\\

\end{tabular}
\end{ruledtabular}
\end{table}

We substitute the quasipotential (\ref{eq:pot}) in the quasipotential equation (\ref{quas}) and solve the resulting differential equation numerically. The calculated masses $M$ of the neutral $QQ'\bar Q\bar Q'$ tetraquarks  composed of the heavy diquark ($QQ'$, $Q=b,c$), and heavy antidiquark ($\bar Q\bar Q'$) are given in
Table~\ref{tab:QQmass}. The masses of the charged heavy $QQ'\bar Q\bar Q'$ tetraquarks are presented in Table~\ref{tab:QQQQ'mass}. In these tables we give the values of the
lowest thresholds $T$ for decays into two corresponding
heavy mesons [$(Q\bar Q')$], which were calculated
using the measured masses of these mesons \cite{pdg}. We also show
values of the difference of the tetraquark and threshold masses,
$\Delta=M-T$. If this quantity is negative, then the tetraquark
lies below the threshold of the fall-apart decay into two mesons  and thus should be a narrow state. The states with small positive values of $\Delta$
could be also observed as resonances, since their decay rates will
be suppressed by the phase space. All other states are expected to
be broad and thus  difficult to observe. 

From these tables we see that the predicted masses of almost all $QQ\bar
Q\bar Q$ tetraquarks lie significantly higher than the thresholds of
the fall-apart decays to the lowest allowed two quarkonium states. All
these states should be broad, since they can decay to corresponding
quarkonium states through quark and antiquark rearrangements,  and
these decays  are
not suppressed either dynamically or 
kinematically. This conclusion is in accord with the current experimental
data. Indeed, the CMS \cite{cms} and LHCb \cite{lhcb} Collaborations have not
observed narrow beautiful tetraquarks in the $\Upsilon(1S)$-pair
production. Note that the lattice nonrelativistic QCD \cite{lattnrqcd}
calculations did not find a signal for the $bb\bar b\bar b$
tetraquarks below the lowest noninteracting two-bottomonium threshold.
On the other hand the broad structure near the di-$J/\psi$
mass threshold very recently observed by the LHCb \cite{Aaij:2020fnh}
can correspond to the $2^{++}$ state of the $cc\bar c\bar c$
tetraquark, with a mass predicted to be 6367 MeV. The narrow
structure, $X(6900)$ \cite{Aaij:2020fnh}, could be the orbital or radial excitation of
this tetraquark.  Such excited states can be
narrow despite the large phase space since it will be necessary in the
fall-apart process to overcome the suppression either due to the
centrifugal barrier for the orbital excitations or due to the presence
of the nodes in the wave function of the radially excited state.          

\begin{table}
 \caption{Comparison of theoretical predictions for the masses
 of the neutral  $(QQ)(\bar Q \bar Q)$ tetraquarks  composed from the same flavour heavy quarks and antiquarks (in MeV).}
 \label{tab:cm1}
\begin{ruledtabular}
\begin{tabular}{ccccccc}
Reference & \multicolumn{3}{c}{$cc \bar c \bar c$}& \multicolumn{3}{c}{$bb \bar b \bar b$}\\
\cline{2-4} \cline{5-7} & $0^{++}$ & $1^{+-}$ & $2^{++}$ & $0^{++}$ & $1^{+-}$ & $2^{++}$\\
\hline
 \centering{this paper} & 6190 & 6271 & 6367 & 19314 & 19320 & 19330 \\
  \centering{\cite{D12},\cite{blln}} & 5966 & 6051 & 6223  & 18754 & 18808 & 18916 \\
 \centering{\cite{SumR1}} & 6460-6470  & 6370-6510 & 6370-6510 & 18460-18490 & 18320-18540 & 18320-18530 \\
 \centering{\cite{FullHeavy2017,Karliner:2020dta}} & $6192 \pm 25$ & &$6429\pm25$ & $18826 \pm 25$ & & $18956\pm25$\\
  \centering{\cite{SumR2,E18}} & $5990 \pm 80$ & $6050 \pm 80$ & $6090 \pm 80$   & $18840 \pm 90$ & $18840 \pm 90$ & $18850 \pm 90$\\
  \centering{\cite{Chiral}} & 6797 & 6899 & 6956 & 20155 & 20212 & 20243\\
  \centering{\cite{FullHeavy2018}} &$ <6140$ & &  & 18750 & & \\
  \centering{\cite{FullCharm2017,E19}} & 5969 & 6021 & 6115  
\\
  \centering{\cite{FullHeavy2019,liu:2020eha}} & 6487 & 6500 & 6524 & 19322 & 19329 & 19341 \\
  \centering{\cite{FullHeavy2019sec}} & 5883 & 6120 & 6246 & 18748 & 18828 & 18900\\
 \centering{\cite{FullBeauty2019}}&&& & $18690 \pm 30$ & &\\
 \centering{\cite{WLZ}}&6425&6425&6432&19247 &19247  &19249 \\
 \centering{\cite{DCP}}&6407&6463&6486&19329&19373&19387\\
  \centering{\cite{Chen}}&&&&19178&19226&19236\\
  \centering{\cite{Jin:2020jfc}}&6314&6375&6407&19237&19264&19279\\
  \centering{\cite{Lu:2020cns}}&6542&6515&6543&19255&19251&19262\\
  \end{tabular}
\end{ruledtabular}
\end{table}

\begin{table}
 \caption{Comparison of theoretical predictions for the masses
 of the  $(cb)(\bar c \bar b)$ tetraquarks (in MeV).}
 \label{tab:cm2}
\begin{ruledtabular}
\begin{tabular}{ccccccc}
\centering{Reference} & \multicolumn{3}{c}{{$A \bar A$}} & \multicolumn{2}{c}{{$\frac{1}{\sqrt{2}}(A \bar S \pm S \bar A)$}} & \multicolumn{1}{c}{{$S \bar S$}}
\\[1ex]
\cline{2-4} \cline{5-6} \cline{7-7}
 & \centering{$0^{++}$} & \centering{$1^{+-}$} & \centering{$2^{++}$} & \centering{$1^{++}$} & \centering{$1^{+-}$} & \multicolumn{1}{c}{$0^{++}$}
\\
\hline
\centering{this paper} & 12813 & 12826 & 12849 & 12831 & 12831 & 12824
\\
  \centering{\cite{D12}} &12359 & 12424  & 12566 & 12485 &12488 & 12471
\\
  \centering{\cite{Chiral}} & 13483 & 13520 & 13590 & 13510 & 13592 & 13553
\\
  \centering{\cite{FullHeavy2018}} & $<12620$ & & & & &
\\
  \centering{\cite{FullHeavy2019}} & 13035  & 13047 & 13070 & 13056 & 13052 & 13050
\\
  \centering{\cite{FullHeavy2019sec}} & 12374 & 12491 & 12576 &12533 & 12533 & 12521
\\
\centering{\cite{DCP}}&12829&12881&12925&\\
\centering{\cite{Chen2}}&12746&12804&12809&&12776
\end{tabular}
\end{ruledtabular}
\end{table}

\begin{table}
 \caption{Comparison of theoretical predictions for the masses
 of the charged $(QQ)(\bar Q\bar Q')$ tetraquarks (in MeV).}
 \label{tab:cm3}
\begin{ruledtabular}
\begin{tabular}{ccccc}
  &  \multicolumn{3}{c}{$A \bar A$} & \multicolumn{1}{c}{$A \bar S, S \bar A$}
\\[1ex]
\cline{2-4} \cline{5-5} 
Reference & $0^{+}$ & $1^{+}$ & $2^{+}$ & $1^{+}$\\[1ex]
\hline
&\multicolumn{4}{c}{$cc \bar c \bar b$, $cb \bar c \bar c$}
\\[1ex] 
This paper & 9572 & 9602 & 9647 &  9619 \\
  \cite{Chiral} & 10144 & 10282 & 10273 & 10174  
\\
  \cite{FullHeavy2018} & $<9390$ & & &  
\\
  \cite{FullHeavy2019} & 9740 & 9749 & 9768 & 9746 
  \\
 \centering{\cite{DCP}}& 9670&9683&9732\\
\\[1ex]
& \multicolumn{4}{c}{$bb \bar c \bar b$, $cb \bar b \bar b$ }
\\[1ex]
 This paper & 16109 & 16117 & 16132 & 16117 \\
  \cite{Chiral} & 16823 & 16840 & 16917 & 16915
\\
  \cite{FullHeavy2018} & $<15770$ & & & 
\\
  \cite{FullHeavy2019} & 16158 & 16164 & 16176 & 16157 
\\
\centering{\cite{DCP}}&16126&16130&16182\\

\end{tabular}
\end{ruledtabular}
\end{table}

\begin{table}
 \caption{Comparison of theoretical predictions for the masses
 of the $cc \bar b \bar b$, $bb \bar c \bar c$ tetraquarks (in MeV).}
 \label{tab:cm4}
\begin{ruledtabular}
\begin{tabular}{cccc}
  &  \multicolumn{3}{c}{$A \bar A$} \\
\cline{2-4} 
Reference & $0^{+}$ & $1^{+}$ & $2^{+}$\\
\hline
This paper & 12846 & 12859 & 12883 
\\
  \cite{Chiral} & 13496 & 13560 & 13595 
\\
  \cite{FullHeavy2018} & $<12580$ & & 
\\
  \cite{FullHeavy2019} & 12953 & 12960 & 12972 
\\
  \cite{FullHeavy2019sec} & 12445 & 12536 & 12614  
\\
\centering{\cite{WLZ}}&12866&12864&12868
\\
\centering{\cite{DCP}}&12906&12945&12960
\\
\centering{\cite{Chen2}}&12892&12898&12905\\
\end{tabular}
\end{ruledtabular}
\end{table}

In Tables~\ref{tab:cm1}-\ref{tab:cm4} we compare our predictions for
the masses of $QQ\bar Q\bar Q$ tetraquarks with the results of
previous calculations
\cite{D12,blln,SumR1,FullHeavy2017,Karliner:2020dta,SumR2,E18,Chiral,FullHeavy2018,FullCharm2017,E19,FullHeavy2019,liu:2020eha,FullHeavy2019sec,FullBeauty2019,
  WLZ,DCP,Chen,Jin:2020jfc,Lu:2020cns,Chen2}. The nonrelativistic
quark model and diquark-antidiquark structure of tetraquarks was employed in
Ref.~\cite{D12,blln}, while the authors of
Refs.~\cite{FullHeavy2017,Karliner:2020dta} used for the calculations the string-junction
picture and the constituent diquark-antidiquark model. References
\cite{SumR1,SumR2,E18} present results obtained in different versions
  of the QCD sum rules. A simple constituent quark model with the
  color-magnetic interaction was applied in Ref.~\cite{Chiral}. The
  relativized diquark-antidiquark model and variational method with
  harmonic oscillator trial wave functions were employed in
  Refs.~\cite{FullHeavy2018,FullHeavy2019sec}, mass inequality
  relations among tetraquarks and heavy quarkonia were also obtained
  \cite{FullHeavy2018}. Different versions of the nonrelativistic
  quark model and   diquark-antidiquark picture were used in
  Refs.~\cite{FullCharm2017,E19,FullHeavy2019,liu:2020eha,WLZ}. The diffusion
  Monte Carlo method was applied to solve the nonrelativistic
  four-body problem for the $bb\bar b\bar b$ tetraquark in
  Ref.~\cite{FullBeauty2019}. In Ref.~\cite{DCP} the multiquark color
  flux-tube model was employed. The meson-meson and
  diquark-antidiquark structures were considered: in the
  nonrelativistic chiral quark model using the Gaussian expansion method
  in Refs.~\cite{Chen,Chen2}; in the nonrelativistic quark
  delocaliztion color screening model using the resonating group method
  for bound states \cite{Jin:2020jfc}, and in the extended relativized quark
  model using variational approach with Gaussian wave functions in
Ref.~\cite{Lu:2020cns}.  The diquark-antidiquark picture with the potential
taken from lattice calculations was studied in
Ref.~\cite{Giron:2020wpx} and the
simplifying dynamical assumptions were investigated:  whether color-sextet
diquark couplings are suppressed, and whether spin couplings between the
diquark and antidiquark are suppressed. Note that we and authors of
Refs.~\cite{D12,blln,FullHeavy2017,Karliner:2020dta,FullHeavy2018,
  FullCharm2017,E19,FullHeavy2019sec} consider diquarks and antidiquarks only in the
color triplet and antitriplet color states, while the color sextet and
antisextet configurations and their mixing are accounted for in
Refs.~\cite{Chiral,FullHeavy2019,liu:2020eha,WLZ,DCP,Chen,Jin:2020jfc,Lu:2020cns,Chen2}. In
most of the previous calculations diquarks and antidiquarks were
considered to be point-like. Our calculation shows that the account of
the diquark structure (size)  weakens the Coulomb-like one-gluon exchange potential, thus increasing tetraquark masses and
reducing spin-spin splittings. We can see from
Tables~\ref{tab:cm1}-\ref{tab:cm4} that there are significant
disagreements between different theoretical approaches. Indeed,
Refs.~\cite{D12,blln,SumR2,E18,FullCharm2017,E19,FullHeavy2019sec,FullBeauty2019} predict heavy
tetraquark masses below or slightly above the thresholds of the decays
to two quarkonia and, thus, stable or significantly suppressed against
fall-apart decays with a very narrow decay width. On the other hand
our model and other approaches predict such tetraquark masses
significantly above these thresholds and, thus, they can be observed only as broad
resonances.  Note that the arguments that these tetraquarks should be unbound
were also given on the basis of the hyperspherical harmonic
expansion \cite{vbv}, the string dynamics \cite{rvv}, and the Hall–Post
inequalities \cite{rvv2}.

\section{Conclusions}
\label{sec:concl}

We calculated the masses of ground-state tetraquarks composed only of
heavy ($b$ or/and $c$) quarks and antiquarks in the framework of the
diquark-antidiquark picture and relativistic quark model based on the
quasipotential approach. It was assumed that two heavy quarks and two
heavy antiquarks will form a doubly-heavy diquark and antidiquark,
respectively. The dynamics of quarks in the diquark is governed by the
relativistic $QQ$ quasipotential which is one half of the $Q\bar Q$
potential in the heavy quarkonium. Masses and wave functions of
diquarks were calculated by the numerical solution of the
quasipotential equation. The obtained diquark wave functions were used
for the evaluation of the form factors of the diquark-gluon
interaction $F(r)$. Then the $QQ\bar Q\bar Q$ tetraquark was
considered as a bound diquark-antidiquark system. It was assumed that
diquarks and antidiquarks interact as a whole. Constructing the
quasipotential of the $d-\bar d$ interaction the same
assumptions about the structure of the long-range confining interaction
were used with the correction to the integer spin of the diquark. In
the potential of the one-gluon exchange between the diquark and
antidiquark the  form factors $F(r)$ of the
diquark-gluon interaction were introduced. They are expressed as the
overlap integrals of the diquark wave functions and take into account the
internal structure (finite size) of the diquarks and
antidiquarks. These form factors
significantly weaken the Coulomb-like potential, thus increasing thr
masses of the tetraquarks and reducing spin splittings. This effect is
especially pronounced for the $bb\bar b \bar b$ tetraquarks since they
have a larger Coulomb contribution due to their  smaller size. 
Note that the approaches with a
point-like diquark substantially underestimate the mass of the doubly
charmed baryon $\Xi_{cc}$, while our model correctly predicted its
mass \cite{efgm} long before its experimental discovery.

It was found that the predicted masses of all ground-state $QQ\bar Q\bar Q$
tetraquarks are above the thresholds for decays into two heavy ($Q\bar Q$)
mesons. Therefore they should rapidly fall apart into the two lowest
allowed quarkonium states. Such decays proceed through quark
rearrangements and are not suppressed dynamically or
kinematically. These states should be broad  and are thus difficult to
observe experimentally. The states, with masses predicted to be less than
200 MeV higher than the lowest allowed thresholds, are  the $1^{+-}$ and  $2^{++}$
states of the  $cc\bar c\bar c$ tetraquark. They have the smallest
phase space for the decay to two charmonium states. The former one decays
mainly to
$\eta_c J/\psi$ while the latter one decays to $J/\psi J/\psi$.  The $2^{++}$
$cc\bar c\bar c$ state with the predicted mass 6367 MeV can correspond
to the broad structure  recently observed by the LHCb Collaboration
\cite{Aaij:2020fnh} in the mass spectrum of  $J/\psi$-pairs produced
in proton-proton collisions.  On the other hand all ground-state
$bb\bar b\bar b$ tetraquarks have masses significantly (400--500 MeV)
higher than corresponding thresholds, and thus should be very
broad. This agrees well with the absence of the narrow beautiful
tetraquarks in the $\Upsilon$-pair production reported by the CMS
\cite{cms} and LHCb \cite{lhcb} Collaborations. 

\acknowledgments
We are grateful to A. Berezhnoy and D. Ebert  for support and valuable
discussions.

\bibliography{tetrQQQQ}

\end{document}